\documentstyle[aps,epsf,twocolumn]{revtex}
\begin{document}
\title
{
Casimir effect in dielectrics: Surface area contribution
}
\author{
C. Molina--Par\'{\i}s$^{+}$ and Matt Visser$^{++}$
}
\address{
$^{+}$Theoretical Division, Los Alamos National Laboratory,
Los Alamos, New Mexico 87545\\
$^{++}$Physics Department, Washington University,
St. Louis, Missouri 63130-4899\\
}
\date{hep-th/9707073; 8 July 1997}

\maketitle

\section*{Abstract}
{\small In this paper we take a deeper look at the technically
elementary but physically robust viewpoint in which the Casimir energy
in dielectric media is interpreted as the change in the total zero
point energy of the electromagnetic vacuum summed over all states.
Extending results presented in previous papers [hep-th/9609195;
hep-th/9702007] we approximate the sum over states by an integral over
the density of states {\em including finite volume corrections}.  For
an arbitrarily-shaped finite dielectric, the first finite-volume
correction to the density of states is shown to be proportional to the
surface area of the dielectric interface and is explicitly evaluated
as a function of the permeability and permitivity.  Since these
calculations are founded in an elementary and straightforward way on
the underlying physics of the Casimir effect they serve as an
important consistency check on field-theoretic calculations.  As a
concrete example we discuss Schwinger's suggestion that the Casimir
effect might be the underlying physical basis behind {\em
sonoluminescence}. The recent controversy concerning the relative
importance of volume and surface contributions is discussed. For
sufficiently large bubbles the volume effect is always
dominant. Furthermore we can explicitly calculate the surface area
contribution as a function of refractive index.

\medskip

PACS: 12.20.Ds; 77.22.Ch;  78.60.Mq
}

\def\Re{{\rm Re}}
\def\Im{{\rm Im}}
\def\small{}
\section{Introduction}

The Casimir effect in dielectrics is the subject of intense
on-going interest. While there is no doubt that the effect is real,
certain suggested applications are somewhat controversial. For
instance: Schwinger has suggested that the Casimir effect might be the
underlying physics behind sonoluminescence
\cite{Schwinger0,Schwinger1,Schwinger2}, while Carlson, Goldman, and
P\'erez--Mercader have suggested possible applications to Gamma Ray
Bursts \cite{GRB}. More generally, the Casimir energy has sometimes
been invoked as a possible driving mechanism for ultra-high-energy
astrophysical processes such as quasars. We feel that all aspects of
the discussion could benefit from the improved understanding of the
basic physics we provide in this paper.

Historically, the techniques used to investigate the Casimir effect
were typically a varied mixture of Schwinger's source theory, explicit
calculations of electromagnetic Green functions (seasoned with
time-splitting regularization), and sometimes, more physically based
regulator schemes that take advantage of the analyticity properties of
the frequency dependent refractive index. A key early paper is that by
Schwinger, de Raad and Milton~\cite{Schwinger-deRaad-Milton}.

Schwinger's most developed point of view can be gleaned from the
series of papers he recently wrote wherein he explored the possible
relevance of the Casimir effect to
sonoluminescence~\cite{Schwinger0,Schwinger1,Schwinger2}. For the
evolution of his views on this subject
see~\cite{Schwinger3,Schwinger4,Schwinger5,Schwinger6}.

Schwinger found~\cite{Schwinger0} that (for each polarization state)
the {\em ``dielectric energy, relative to the zero energy of the
vacuum, [is given] by}

\begin{equation}
\label{E-Schwinger-0}
E = - V \int \frac{d^3\vec{k}}{(2 \pi)^3} \frac{1}{2} 
     \, [\hbar c] \, k  
     \left( 1 - \frac{1}{\sqrt{\epsilon}} \right).\hbox{\em ''}
\end{equation}

\noindent
This result can be interpreted in a straightforward manner as the
integral of the {\em difference} in dispersion relations over the
density of states~\cite{CMPV-short,CMPV-long}.

In addition to Schwinger's bulk volume term, calculations by Milton
{\em et al.}~\cite{Milton80,Milton95,Milton96} indicate the existence
of a surface correction.  For a dilute (that is, $\epsilon \approx 1$)
spherical intrusion of radius $R$ and dielectric constant $\epsilon_1$
in a dilute dielectric medium of dielectric constant $\epsilon_2$
($\epsilon_2 \approx 1$), with the eigenmode sum regulated by
time-splitting, the surface contribution is equivalent to\footnote{See
equation (51) of~\cite{Milton95}, equation (7.5) of~\cite{Milton96},
or the equivalent equation (41) of~\cite{Milton80}.  Those
calculations only deal with spherical dielectric balls with frequency
independent dielectric properties, and use an explicit time-splitting
regularization. The numerical coefficient in this surface term is
regularization dependent and it does not appear to be possible to
relate its absolute normalization to the number we will calculate
using Schwinger's wave-number cutoff.}

\begin{equation}
\label{E-Milton-0}
E_{surface} 
\approx 
- {1\over4}  \hbar c (\epsilon_1 - \epsilon_2 )^2 R^2 
{1\over (c\tau)^3}.
\end{equation}

A controversy has recently arisen over whether or not Schwinger's
volume term should be retained, and whether or not the surface term is
the leading term in the Casimir energy~\cite{Milton95,Milton96}.  We
have shown elsewhere~\cite{CMPV-short,CMPV-long} that the presence of
the volume term is generic, and have (among other arguments) adduced
reasons based on density-of-states calculations to bolster Schwinger's
calculation. In this paper we shall pursue this matter further and
shall extract as much physics as possible from these density-of-states
calculations.

The discussion, though elementary from a technical perspective, is
quite sufficient to give the most important dominant contributions to
the Casimir energy. These results serve as an important consistency
check on more sophisticated field-theoretic calculations.

Furthermore, the present analysis extends Schwinger's result by
verifying that generically surface terms do in fact show up, but as
{\em sub-dominant} corrections to the dominant volume contribution.
General arguments of this type are particularly useful because they
allow us to study {\em arbitrary} shapes and not be limited by
requirements of spherical symmetry.

We first discuss some general properties of the bulk volume term,
noting in particular the dependence upon a physically meaningful
ultraviolet cutoff, and then turn to the issue of finite volume
effects. While finite volume effects in conductors (or more precisely,
for Dirichlet, Neumann, and Robin {\em boundary} conditions) are well
understood, the analogous problem for dielectric {\em junction}
conditions (or even acoustic junction conditions) is considerably less
clear cut. We attack the problem of finite volume effects in the
presence of junction conditions via an extension of the Balian--Bloch
analysis for boundary
conditions~\cite{Balian-Bloch,Balian-Bloch-71}. We show that the
presence of a dielectric interface modifies the density of states by a
term proportional to the surface area of the interface and calculate
the proportionality constant as an explicit function of the dielectric
permitivity and permeability. (For the related, and simpler, acoustic
interface the change in density of states is related to the physical
fluid densities on the two sides of the interface.)

Finally, we apply this formalism to the estimation of the
(electromagnetic) Casimir energy in generic dielectrics. We show that
for dielectric bubbles large compared to the cutoff wavelength the
volume term is dominant. We point out that the numerical value of the
net Casimir energy is strongly dependent on the details of the high
frequency cutoff. Within the context of sonoluminescence this
high-frequency sensitivity might explain the fact that small
admixtures of gas in the bubble undergoing sonoluminescence can have
large effects on the total energy radiated: a small resonance in the
medium-frequency behaviour of the refractive index can be magnified by
phase space effects, and lead to dramatic changes in the total energy
budget.

We mention in passing that there will also be an acoustic
Casimir energy associated with the phonon modes~\cite{Visser}. The
acoustic Casimir energy (while always present) is numerically
negligible in comparison to the electromagnetic effect being
suppressed by a factor of $(\hbox{speed of sound}/\hbox{speed of
light})$.

\section{The density of states: Bulk term}

The physics underlying the Casimir effect is that every eigenmode of
the photon field has zero point energy $E_n = (1/2)\hbar \omega_n$;
the Casimir energy is the difference in zero point energies
between any two well defined physical situations

\begin{equation}
\label{E-Casimir-primitive}
E_{\small Casimir}(A \mid B) = 
\sum_n {1\over2} \hbar \left[\omega_n(A) - \omega_n(B)\right].
\end{equation}

\noindent
We always need a regulator to make sense of this energy
difference, though in many cases of physical interest (such as
dielectrics) the physics of the problem will automatically regulate
the difference for us and make the results finite. Adding over
all eigenmodes is prohibitively difficult, so it is in general
more productive to replace the sum over states by an integral over
the density of states.

Suppose we have a finite volume $V$ of some bulk dielectric in
which the dispersion relation for photons is given by some function
$\omega_1(k)$, which describes the photon frequency as a function
of the wave-number (three-momentum) $k$. Suppose this dielectric
to be embedded in an infinite background with different dielectric
properties described by a different dispersion relation $\omega_2(k)$.
We regulate infra-red divergences by putting
the whole universe in a box of finite volume $V_\infty$, and 
calculate the bulk contribution to the total zero-point energy of
the electromagnetic field by summing the photon energies
over all momenta (and polarizations), using the usual and elementary
density of states: $\hbox{[Volume]} \, d^3 \vec k /(2\pi)^3$.  (In
the next section we shall look at finite-volume corrections to this
density of states.)

Including photon modes both inside and outside the dielectric body the
energy of the system is

\begin{eqnarray}
\label{E-embedded-body}
E_{\small embedded-body} 
&=& 
2 V \int {d^3 \vec k\over(2\pi)^3} \; {1\over2} \hbar \;
\omega_1(k)
\nonumber\\
&+& 
2 (V_\infty - V) 
\int {d^3 \vec k\over(2\pi)^3} \; {1\over2} \hbar \; \omega_2(k).
\end{eqnarray}

\noindent
Note that outside the dielectric body the photon dispersion relation
is that of the embedding dielectric $\omega_2(k)$.  Note also that we
shall always use the subscript $2$ to refer to the region {\em
outside} the embedded body, and shall use the subscript $1$ to refer
to the region {\em inside}.

If the embedded body is removed, and the hole simply filled in with
the embedding medium, we can calculate the total zero-point
energy as

\begin{equation}
\label{E-homogeneous}
E_{\small homogeneous} 
= 2 V_\infty \int {d^3 \vec k\over(2\pi)^3} \; 
{1\over2} \hbar \; \omega_2(k).
\end{equation}

\noindent
We {\em define} the Casimir energy by subtracting these two
zero-point energies~\cite{CMPV-short,CMPV-long}

\begin{eqnarray}
\label{E-Casimir}
E_{\small Casimir} 
&\equiv&  
E_{\small embedded-body} - E_{\small homogeneous}
\nonumber \\
&=&
2 V \int {d^3 \vec  k\over(2\pi)^3} \; 
{1\over2} \hbar \; [\omega_1(k) - \omega_2(k)].
\end{eqnarray}

\noindent
The physical import of this definition is clear: The Casimir energy
is defined as the {\em change} in the zero-point energy due to a
change in the medium.

Note also that the physical meaning of the zero of energy is clear:
the zero of energy is here taken to be that corresponding to a
homogeneous dielectric with dispersion relation $\omega_2(k)$. 

To be obtuse, we could use a different zero for the energy --- this
makes no difference as long as we keep the same zero throughout any
particular calculation. For instance, the zero-point energy of the
Minkowski vacuum is

\begin{equation}
\label{E-Minkowski}
E_{\small Minkowski} 
= 2 V_\infty \int {d^3 \vec k\over(2\pi)^3} 
\; {1\over2} \hbar c k.
\end{equation}

\noindent
Thus an alternative  {\em definition} for the Casimir energy is then

\begin{eqnarray}
\label{E-Casimir-alternative}
E_{\small Casimir}^{\small alternative} 
&\equiv&  E_{\small embedded-body} - E_{\small Minkowski}
\nonumber \\
&=&
2 V \int {d^3 \vec  k\over(2\pi)^3} \; 
{1\over2} \hbar \; [\omega_1(k) - c k]
\nonumber \\
&+&
2 (V_\infty - V) 
\int {d^3 \vec  k\over(2\pi)^3} \; 
{1\over2} \hbar \; [\omega_2(k) - c k].
\end{eqnarray}

\noindent
For this alternative definition, the zero of energy is clearly the
Minkowski vacuum. As long as you stick with one fixed definition
throughout the calculation, or better yet, calculate Casimir energy
differences directly, quibbling about the zero of energy does not
matter. (Of course, if you change the zero of energy in the middle 
of the calculation the answers will be meaningless.)

{From} the general considerations in~\cite{CMPV-short,CMPV-long} we
know that the integrand must go to zero at large wave-number, and in
fact, for any pair of real physical dielectrics the integrand must go
to zero sufficiently rapidly to make the integral converge.

An integration by parts yields

\begin{eqnarray}
E_{\small Casimir}  
&=&
{ V \hbar \over 6 \pi^2 } \int_0^{\infty} 
d(k^3)\; [\omega_1(k) - \omega_2(k)]
\nonumber\\
&=&
{ V \hbar \over 6 \pi^2 } 
\Bigg[ \left.\left(k^3\; 
     [\omega_1(k) - \omega_2(k)]\right)\right|_0^\infty  
\nonumber\\
&& \qquad -  \int_0^{\infty} [d\omega_1(k) - d\omega_2(k)] k^3 \Bigg].
\end{eqnarray}

\noindent
The boundary term vanishes because of the asymptotic behaviour 
of the $\omega_i(k)$. The substitution $k = \omega_i(k) n_i$ 
then yields

\begin{equation}
E_{\small Casimir} = + {V \hbar \over 6 \pi^2 c^3} 
\int_0^\infty \omega^3 [n_2^3(\omega)-n_1^3(\omega)] d\omega.
\end{equation}

\noindent
While the difference between the refractive indices in the above
expression goes to zero sufficiently rapidly to make the integral
converge, it must be noted that the prefactor of $\omega^3$ implies
that the net Casimir energy will be relatively sensitive to the
high frequency behaviour of the refractive indices.

If the Casimir effect ultimately proves to be the correct physical
explanation for sonoluminescence, this sensitivity to the details of
the refractive index might plausibly explain why sonoluminescence is
sensitive to the admixture of small trace gases into the bubble.  (Of
course the present calculation is static, but the energy calculated in
this way will be the maximum energy that could possibly be released in
a more realistic dynamical calculation.) To make this qualitative
statement quantitative we would need a detailed model for the
refractive index as a function of frequency---a task that is beyond
the scope of this paper. 

\section{The density of states: Finite-volume effects}
\subsection{Generalities}

We now look at the contribution arising from sub-dominant finite-volume
corrections to the density of states. The key point here is that
the existence of finite-volume terms proportional to the surface
area of the dielectric is a {\em generic} result.  The fact that
previous calculations~\cite{Milton80,Milton95,Milton96} encountered
a surface tension term proportional to $(\hbox{cutoff})^3$ is hereby
explained on general physical grounds without recourse to special
function theory.

We must notice at this stage that the dominant contribution to the
Casimir energy is proportional to volume, as the canonical bulk
expression for the density of states is proportional to the volume:
$\hbox{[Volume]}\, d^3 \vec k/(2\pi)^3$.  It is reasonably well-known,
though perhaps not so elementary, that for fields subject to {\em
boundary} conditions (Dirichlet, Neumann, Robin) the density of states
is in general modified by finite volume effects. In this paper we wish
to extend these ideas to fields subject to {\em junction} conditions
(acoustic, dielectric).

For boundary conditions the general result is

\begin{equation}
\label{E-dos}
\sum_n \sim V \int { d^3 \vec k\over (2\pi)^3} +
            S \int \xi { d^3 \vec k \over (2\pi)^3 k } + 
            \cdots
\end{equation}

\noindent
These are the first two terms in an asymptotic expansion in $1/k$.
For Dirichlet, Neumann, and Robin boundary conditions the coefficients
can be related directly to the known asymptotic behaviour of the Heat
Kernel---they are simply the Seeley--DeWitt coefficients in disguise
and can be obtained, for instance, by suitably transforming the results
presented in the monograph by Gilkey~\cite{Gilkey}.

There are additional terms in this expansion, proportional to the
various monomials appearing in the general formulae for the higher
Seeley--DeWitt coefficients, but we do not further address this
issue here except to point out that the next term is proportional
to the integral of the trace of the extrinsic curvature
over the boundary.

An elementary discussion of the general existence of such terms can be
found in the textbook by Pathria~\cite{Pathria}, while a more
extensive treatment can be found in the papers by Balian and
Bloch~\cite{Balian-Bloch,Balian-Bloch-71}.

For Dirichlet, Neumann, and Robin boundary conditions the
dimensionless variable $\xi$ is a known function of the boundary
conditions imposed.

If we let $N(k)$ denote the number of eigenmodes with
wave-number less than $k$, then from the above we can write

\begin{equation}
N(k) \sim {1\over2\pi^2}
\left({1\over3} V k^3 + {1\over2} \xi S k^2 + O[k] \right).
\end{equation}

We shall now perform the analogous analysis for junction conditions,
adapting the Balian--Bloch formalism as needed. Our formalism is
applicable to both boundary conditions and junction conditions. For
clarity, and to aid in consistency checking, we carry out brief
parallel computations for the boundary condition case.

\subsection{Scalar field}

We start for simplicity with a scalar, rather than electromagnetic,
field. We are interested in the following eigenvalue problem

\begin{equation}
\Delta \phi + k^2 \phi = 0; \qquad B[\phi] = 0.
\end{equation}

\noindent
Here $B[\phi]$ denotes the boundary conditions imposed. Common
boundary conditions are tabulated below.

\vskip 0.25 cm

\noindent
{\bf Dirichlet boundary conditions:}\\ 
($\phi = 0$ on the boundary) 

\begin{equation}
\label{E-Dirichlet}
\xi = -\pi/4.
\end{equation}

\noindent
{\bf  Neumann boundary conditions:}\\ 
($\partial_n\phi = 0$ on the boundary; where $\partial_n$ denotes
the normal derivative)

\begin{equation}
\label{E--Neumann}
\xi = +\pi/4.
\end{equation}

\noindent
{\bf Robin boundary conditions:}\\ 
($\partial_n\phi = \kappa \phi$ on the boundary; $\kappa$ real) 

\begin{equation}
\label{E-Robin}
\xi = +\pi/4.
\end{equation}

\noindent
{\bf Surface damped boundary conditions:}\\ 
($\partial_n\phi = k \kappa \phi$ on the boundary; $\kappa$ real;
note that the eigenvalue is now explicitly present in the boundary
condition as well as in the differential equation)

\begin{equation}
\label{E-damped}
\xi 
= {\pi\over4} - 
  {1\over2} \Im\left[\ln\left({1+i\kappa\over1-i\kappa}\right)\right] 
= {\pi\over4} - \arctan(\kappa).
\end{equation}

These results can be read off, for instance, from the paper by
Balian and Bloch~\cite{Balian-Bloch}. 

Comparing the Robin and surface damped boundary conditions, it might
naively be tempting to write

\begin{equation}
\xi_{Robin}(\kappa) = \xi_{damped}(\kappa/k).
\end{equation}

\noindent
However in the present context---an asymptotic expansion in
$1/k$---such an expression is meaningless.  The best we can do is
to say that

\begin{equation}
\xi_{Robin}(\kappa) = \lim_{k\to\infty} \xi_{damped}(\kappa/k).
\end{equation}

\noindent
Thus for Robin boundary conditions we keep only the dominant
$k\to\infty$ piece of the Balian--Bloch result.

On the other hand, in the surface damped boundary condition (because
of the explicit factor of $k$ appearing in this boundary condition) it
{\em is} meaningful to keep the inverse tangent term  of the
Balian--Bloch result in our expression for $\xi$.  (As a consistency
check, these coefficients are also calculated as special cases of the
general formalism we shall develop below.)

\vskip 0.25 cm

\noindent
{\bf Acoustic junction conditions:}\\ 
We are  ultimately interested in {\em junction conditions}, rather
than {\em boundary conditions}. For definiteness, we can think of
an acoustic junction, wherein an acoustic wave propagates across
some fluid interface: say a bubble of some dense fluid embedded in
a lighter fluid.  In terms of the densities of the fluids, ($\rho_1$,
$\rho_2$), and the velocity potentials, ($\phi_1$, $\phi_2$), the
acoustic junction conditions are

\begin{equation}
\label{E-junction-condition-1}
\rho_1 \phi_1 = \rho_2 \phi_2,
\end{equation}

\begin{equation}
\label{E-junction-condition-2}
\partial_n \phi_1 = \partial_n \phi_2.
\end{equation}

\noindent
(See \cite[page 24]{DeSanto} or \cite[page 81]{DeSanto}. These two
conditions represent, respectively, the continuity of the pressure and
the normal component of the velocity at the interface.) We must point
out at this stage that the change in propagation speed and/or density
causes a certain amount of reflection and refraction, which then
changes the density of states in the fluid both {\em inside} and {\em
outside} the bubble ({\em i.e.} on both sides of the interface)
according to the general scheme

\begin{eqnarray}
\sum_{inside} 
&\sim& 
V \int { d^3 \vec k\over (2\pi)^3} +
S \int \xi_{\small in} { d^3 \vec k \over (2\pi)^3 k} + 
\cdots
\\
\sum_{outside} 
&\sim& 
(V_\infty-V) \int { d^3 \vec k\over (2\pi)^3} +
S \int \xi_{\small out} { d^3 \vec k \over (2\pi)^3 k} + 
\cdots
\end{eqnarray} 

\noindent
For the case of acoustic junction conditions, the dimensionless
variables $\xi_{\small out/in}$ have not yet been calculated.  We
present the calculation below, for now merely quoting the final
result:

\begin{equation}
\xi_{\small out} (\rho_1,\rho_2)
={\pi\over4} \left[ {\rho_1-\rho_2\over\rho_1+\rho_2}\right] =
- \xi_{\small in}(\rho_1,\rho_2).
\end{equation}

\bigskip
\noindent
{\bf Formulation of the problem:}\\
On general grounds, we expect the $\xi$ to be a function of
both the acoustic refractive index (that is, a function of the
relative acoustic velocities), and the relative densities. If $v_0$
is some arbitrary reference speed, we can define the refractive
indices by

\begin{equation}
n_1\equiv k_1 v_0/\omega  
\qquad \hbox{and} \qquad 
n_2\equiv k_2 v_0/\omega,
\end{equation}

\noindent
and further define the relative refractive index by  $n = n_1/n_2$.
(Note in particular that $\omega$ is continuous across the interface,
whereas $k_i$ is not.) It is also useful to define the density
contrast by $\rho = \rho_1/\rho_2$.

In the special case where there is no dispersion, the phase and
group velocities are equal and we simply have

\begin{equation}
n_1=v_0/v_1 \qquad \hbox{and} \qquad n_2=v_0/v_2.
\end{equation}

We know, from first principles, that as $n\to 1$ and $\rho\to1$ the
acoustic boundary becomes indistinguishable, as both fluids have the
same density and refractive index, so we must have

\begin{equation}
\xi_{\small out/in}(n,\rho)\to 0 \qquad 
{\rm as} \qquad n\to 1 \ \ {\rm and} \ \ \rho\to 1.
\end{equation}

\noindent
To calculate $\xi(n,\rho)$ for acoustic junction conditions, we modify
the discussion of Balian and Bloch~\cite[page 407]{Balian-Bloch} to
derive an expression for $\xi(n,\rho)$ in terms of an integral
involving the reflection coefficient $R(\rho,n;\vec k)$.

We start from the result for the density of states in terms of the
time-independent Green function~\cite[equation (II.6), page
409]{Balian-Bloch}. Taking $N(k)$ to be the number of modes with
wave-number less than $k$,  we can construct a suitably smoothed
density of states formally described by the relation

\begin{equation}
\rho_{\small dos}(k) = 
\left[{dN\over dk}\right]_{\small smoothed}.
\end{equation}

\noindent
(Details of the smoothing procedure can be found
in~\cite{Balian-Bloch}.)  Note that we prefer to express the density
of states in terms of the wave-number $k_i$ rather than in terms of
the variable $E=k_i^2$.  See equation (I.3) on page 402
of~\cite{Balian-Bloch}. Thus

\begin{equation}
\rho_{dos}(k) 
\sim {dN\over dk} 
\sim {dE\over dk} {dN\over dE}
\sim 2 k \; \rho^{BB}_{dos}(E).
\end{equation}

\noindent
In terms of the asymptotic expansion of interest

\begin{equation}
\rho_{dos}(k) 
\sim {1\over2\pi^2}  \left( V k^2 + S \xi k + O[1] \right).
\end{equation}

Working on either side of the interface (with $i$ taking on the values
``in'' or ``out'' as appropriate) equation (II.6) on page 409
of~\cite{Balian-Bloch} yields

\begin{equation}
\rho^i_{\small dos}(k_i) = 
{2 k_i\over\pi} \int d^3  \vec x \; 
\lim_{\vec x' \to \vec x} \Im [G(\vec x,\vec x';k_i+i\epsilon)],
\end{equation}

\noindent
where the integration over $x$ now runs only over region $i$
as appropriate.

It is important to realize that the Balian--Bloch formalism is
built up under the assumption that all the  eigenvalues are
real---this constrains the type of problems we can deal with to
loss-free undamped situations.

We are interested in an arbitrary interface, but provided the
interface is smooth, we can locally replace it by its tangent
plane. This approximation is equivalent to neglecting sub-dominant
pieces proportional to the trace of the extrinsic curvature. (If we
were interested in explicitly calculating the next coefficient in the
expansion we would have to locally approximate the surface by its
osculating ellipsoid, as done for the case of boundary conditions by
Balian and Bloch.)

Truncating the expansion at the surface area term, we locally
approximate the interface by a plane interface, located at $z=0$, with
region 2 (the outside) at $z>0$ and region 1 (the inside) at $z<0$.
We are only interested in the diagonal part of the Green function. To
calculate this diagonal part in region $i$ we can assume the source is
also in region $i$ and write the total Green function in this region
as a sum of a {\em direct} and a {\em reflected} contribution.

The direct part of the Green function is responsible for the bulk
contribution to the density of states, while the reflected part of the
Green function gives the surface contribution.  Since, in the tangent
plane approximation, we are dealing with a perfectly flat interface
higher order contributions are explicitly excluded.

The volume contribution has already been calculated
in~\cite{CMPV-short,CMPV-long}, and we are now interested in the extra
piece of the Green function that arises from reflection at the
interface. Using cylindrical coordinates, the contribution to the
Green function due to the reflected wave can be put into the
Sommerfeld representation (an integral over transverse wave-number
$k_t$)

\begin{eqnarray}
&&G^i_{\small reflection}(\vec x,\vec x'; k_i) 
\nonumber\\
&& \qquad = {i\over4\pi} \int_0^\infty 
 R^i(k_i,k_t) J_0(k_t r) 
\nonumber\\
&&\qquad\qquad\times{\exp[i K(k_i,k_t) (z+z') ] \over K(k_i,k_t) }
k_t d k_t.
\end{eqnarray}

\noindent
(See equation (4.2.5) on page 103 of \cite{DeSanto}, with an
appropriate change of notation.) Note that $R(k_i,k_t)$ is the
reflection coefficient. It is a function of the frequency and the
transverse wave-number and will consequently depend on the precise
nature of the boundary conditions imposed.  The Sommerfeld
representation has the interesting feature that it expresses a Green
function, which is related to the behaviour of spherical waves, in
terms of a reflection coefficient defined for plane waves.  Here

\begin{equation}
K(k_i,k_t) = \sqrt{k_i^2 - k_t^2}.
\end{equation}

More explicitly

\begin{equation}
K_{\small out}(k_2,k_t) = \sqrt{k_2^2 - k_t^2}.
\end{equation}

\begin{equation}
K_{\small in}(k_1,k_t) = \sqrt{k_1^2 - k_t^2}.
\end{equation}

\noindent
If we look at the diagonal part of this reflection contribution,
($\vec x = \vec x'$), and note that $J_0(0)=1$ we immediately see

\begin{eqnarray}
&&G^i_{\small reflection}(\vec x,\vec x; k_i) 
\nonumber\\ 
&&\qquad = {i\over4\pi} \int_0^\infty 
{R^i(k_i,k_t) \exp[ 2 i K(k_i,k_t) z ] \over K(k_i,k_t) }
k_t d k_t.
\end{eqnarray}
 
\noindent
(Note that we are calculating what is in field theory parlance an
off-shell Green function. The integration over $k_t$ is an integration
over all off-shell transverse momenta, and this integration is not
to be limited by any on-shell constraint such as $k_t\leq k_i$.)

For the density of states (counting only the appropriate contribution
arising from either side of the interface, that is, $z>0$ or $z<0$)

\begin{eqnarray}
&&\rho^i_{\small reflection}(k_i) 
\nonumber\\
&&\qquad = 
{i k_i\over2\pi^2} S \int_0^\infty d z \;
\Im \Bigg[  \int_0^\infty k_t \;d k_t
\; R^i(k_i+i\epsilon,k_t)
\nonumber\\
&& \qquad \qquad \times {\exp[ 2 i K(k_i+i\epsilon,k_t) z ] 
\over K(k_i+i\epsilon,k_t) } 
\Bigg].
\end{eqnarray}

\noindent
The $z$ integration is trivial. (Because $k_i$ has a small
positive imaginary part, which is inherited by $K(k_i,k_t)$,
we can guarantee convergence of this integral.)

\begin{equation}
\rho^i_{\small reflection}(k_i) = 
-{k_i\over4\pi^2} S \;
 \Im\left[ \; \int_0^\infty 
{R^i(k_i+i\epsilon,k_t)\over K(k_i+i\epsilon,k_t)^2 } 
\; k_t \; d k_t 
\right].
\end{equation}

\noindent
It is useful to define the dimensionless variable
$u=k_t/(k_i+i\epsilon)$, so that $u$ has a small {\em negative}
imaginary part.  We get

\begin{equation}
\label{E-dos-reflection}
\rho^i_{\small reflection}(k_i) = 
-{k_i\over4\pi^2} S \;
 \Im\left[ \; \int_0^\infty 
{R^i(k_i,u-i\epsilon)\over  1-(u-i\epsilon)^2 } 
\; u \; d u 
\right].
\end{equation}

\noindent
If we now take this contribution to the quantity $\rho_{\small dos}$,
and convert to the $\xi^i$ variable as defined in this paper using

\begin{equation}
\xi^i = {2\pi^2\over k S} \; \rho^i_{reflection},
\end{equation}

\noindent we find

\begin{equation}
\xi^i(k_i) = 
-{1\over2}\; 
\Im\left[ \; \int_0^\infty 
{R^i(k_i+i\epsilon,k_t)\over K(k_i+i\epsilon,k_t)^2 }
\; k_t \; d k_t
\right].
\end{equation}
 
\noindent
Equivalently

\begin{equation}
\label{E-general-result-1}
\xi^i(k_i) = 
-{1\over2}\Im\left[ 
\int_0^\infty  {R^i(k_i,u-i\epsilon)\over 1-(u-i\epsilon)^2 } 
\right] u du.
\end{equation}

\noindent
This is our general result for the surface contribution to the density
of states. The surface term is seen to be a suitable average of the
reflection coefficient appropriate to the boundary conditions at hand.
(Note that if we were on-shell, we would interpret $u$ as the sine of
the angle of incidence, and $u$ would then be limited to the range
$u\in[0,1]$. As this is an off-shell computation for the off-shell
Green function, the range of integration goes all the way to infinity
and trying to interpret $u$ as the sine of the angle of incidence only
leads to unnecessary confusion.   Indeed, in calculating this
Green function, we are effectively dealing with a spherical incident
wave, so there are many angles of incidence $\theta_i$.  To identify
$u$ as the sine of {\em the} angle of incidence only makes sense for an
incident plane wave, and is in the present context meaningless.)

The application of this result to specific cases of interest merely
requires us to calculate the relevant reflection coefficients and
perform the integrations.

\bigskip
\noindent
{\bf The integral for standard boundary conditions:}\\
In some well known cases the relevant integrations are straightforward.
For example for Dirichlet, Neumann, and Robin boundary conditions
the reflection coefficients are $-1$, $+1$, and $+1$ respectively,
and integrating out to some large cutoff value of $u$ we have

\begin{eqnarray}
&&\int_0^U  {1\over 1-(u-i\epsilon)^2 } \; u du 
\nonumber\\
&&= 
{1\over2} \int_0^{U^2} {1\over 1 - (x-i\epsilon)} dx
\nonumber\\
&&=
{1\over2} \left. \left[ - \ln\{1-(x-i\epsilon)\} \right] \right|_0^{U^2}
\nonumber\\
&&=
{1\over2} \left( \ln\{1+i\epsilon\} - \ln\{1-U^2+i\epsilon\} \right)
\nonumber\\
&&=
-[i\pi + \ln(U^2-1)]/2
\nonumber\\
&&\approx
-{i\pi\over2}- \ln(U).
\label{E-basic-integral}
\end{eqnarray}

\noindent
Note that the integral itself diverges, though the imaginary part
is both finite and independent of the cutoff.  Taking this imaginary
part gives

\begin{equation}
\xi = \mp {\pi\over4}.
\end{equation}

\noindent
This reproduces the standard results quoted above. 
[Equations (\ref{E-Dirichlet}--\ref{E-Robin}).]

The surface damped boundary condition is a little trickier. In this
case the reflection coefficient can be shown to be

\begin{equation}
R(u) = { \sqrt{1-u^2} - i\kappa  \over \sqrt{1-u^2} + i\kappa}.
\end{equation}

\noindent
(See, for example, equations (3.4.4) and  (3.4.5) on page 87 of
DeSanto~\cite{DeSanto}, and translate to our notation. Note that
an analytic continuation in $\kappa$ is required to turn the surface
impedance boundary condition discussed there into the surface damped
boundary condition discussed here.)

Subtracting and adding 1 to the integrand converts the integral
into that encountered in the previous calculation plus an integral
that is well-behaved at infinity. The relevant integral is again
elementary:

\begin{equation}
\int_0^\infty  {1\over 1-u^2 } 
\left[  {\sqrt{1-u^2} - i\kappa  \over \sqrt{1-u^2} + i\kappa} 
- 1 \right] u du 
= + 2\ln\left(1+i\kappa\right).
\end{equation}

\noindent
Taking the imaginary part of the above reproduces the result
announced in equation (\ref{E-damped}):

\begin{equation}
\xi = {\pi\over4} - \arctan(\kappa).
\end{equation}

\bigskip
\noindent
{\em Checking the above:}

\begin{eqnarray}
&&\int_0^\infty  {1\over 1-(u -i\epsilon)^2} 
\left[  
{\sqrt{1-(u -i\epsilon)^2} - i\kappa  \over 
 \sqrt{1-(u -i\epsilon)^2} + i\kappa} - 1 
\right] u du 
\nonumber\\
&&\qquad=\int_0^\infty  {1\over 1-(u^2 -i\epsilon)} 
\left[  
{\sqrt{1-(u^2 -i\epsilon)} - i\kappa  \over 
 \sqrt{1-(u^2 -i\epsilon)} + i\kappa} - 1 
\right] u du 
\nonumber\\
&&\qquad= \int_0^\infty  {1\over 1-(u^2 -i\epsilon) } 
\left[  
{-2 i\kappa  \over \sqrt{1-(u^2 -i\epsilon)} + i\kappa} 
\right] u d u 
\nonumber\\
&&\qquad= \int_0^\infty  {1\over 1-u'+i\epsilon } 
\left[  
{-i\kappa  \over \sqrt{1-u'+i\epsilon} + i\kappa} 
\right] d u' 
\nonumber\\
&&\qquad= \int_{-1}^\infty  {1\over u'' -i\epsilon} 
\left[  
{i\kappa  \over \sqrt{-u''+i\epsilon} + i\kappa} 
\right] d u'' 
\nonumber\\
&&\qquad= \int_{-1}^\infty  {1\over u''-i\epsilon } 
\left[  
{\kappa  \over \sqrt{u''-i\epsilon} + \kappa} 
\right] d u'' 
\nonumber\\
&&\qquad= \int_{-i}^\infty  {2\over \tilde u -i\epsilon} 
\left[ 
{\kappa  \over \tilde u+\kappa-i\epsilon} \right] 
d\tilde u 
\nonumber\\
&&\qquad= 2 \int_{-i}^\infty 
\left[ 
{1\over \tilde u -i\epsilon} -
{1 \over \tilde u+\kappa-i\epsilon} 
\right] d\tilde u 
\nonumber\\
&&\qquad= 2 
\left[ 
\ln\left( 
{\tilde u -i\epsilon \over  \tilde u+\kappa-i\epsilon} 
\right)
\right]_{-i}^{+\infty}
\nonumber\\
&&\qquad= -2 
\left[ 
\ln\left( 
{-i -i\epsilon \over  -i+ \kappa -i\epsilon} 
\right)
\right]
\nonumber\\
&&\qquad= +2 \ln(1+i\kappa).
\end{eqnarray}

(The original contour was chosen to run underneath the two branch cuts
emanating from $u=-1+i\epsilon$ and $u=+1+i\epsilon$; thus under
the change of variables $u''=\sqrt{u^2-1}$ the branch cut must be
chosen so that the new contour terminates at $-i$ and not at $+i$.)

\bigskip
\noindent
{\bf The integral for acoustic junction conditions:}\\ 
We are finally ready to study the case of interest: acoustic {\em
junction conditions}. The reflection coefficient is now~\cite[equation
(3.1.19), page 82]{DeSanto}

\begin{equation}
R(\rho,n;u) = 
{ \rho \sqrt{1-u^2} - \sqrt{n^2-u^2}  \over 
  \rho \sqrt{1-u^2} + \sqrt{n^2-u^2} }.
\end{equation}

{\em Consistency check I:} Note that $\rho\to+\infty$ gives $R=+1$,
as appropriate for  Neumann and Robin boundary conditions;
$\rho\to0$ gives $R=-1$ as appropriate for the Dirichlet boundary
condition; while $\rho\to\infty$ with $\kappa=-in/\rho$ fixed gives
the surface damped boundary condition.

{\em Consistency check II:} Similarly $n\to+\infty$ gives $R=-1$,
as appropriate for Dirichlet boundary conditions; finally $n\to+i\infty$
gives $R=+1$ as appropriate  for  Neumann and Robin boundary
conditions.

{\em Observation:} The reflection coefficient
exhibits an inversion symmetry as we move from one side of the
interface to the other, this symmetry being inherited by the $\xi$.

\begin{equation}
R_{\small in}(\rho,n;u) = R_{\small out}(1/\rho,1/n;u).
\end{equation}

\noindent
Thus

\begin{equation}
\label{E-inversion}
\xi_{\small in}(\rho,n) = \xi_{\small out}(1/\rho,1/n).
\end{equation}

{\em Calculation:} We are interested in evaluating

\begin{equation}
Q= \Im \left[\int_0^\infty {u du\over1-u^2} \;
{ \rho \sqrt{1-u^2} - \sqrt{n^2-u^2}  \over 
  \rho \sqrt{1-u^2} + \sqrt{n^2-u^2} } \right].
\end{equation}

\noindent
The integrand has a pole at $u=1$ of residue $-1/2$, and branch cuts
emanating from $u=\pm1$ which can be chosen to terminate at $u=\pm
n$. Asymptotically, as $u\to \infty$, the integrand goes as

\begin{equation}
{1\over u} {\rho-1\over\rho +1}.
\end{equation}

\noindent
This is already enough to tell us that the imaginary part of this
integral can be finite if and only if $\rho$ is real. For the acoustic
equations this is actually very sensible physically since it is
meaningless to drive the density complex.  To evaluate this expression
we subtract and add 1 to the integrand, and make use of the integral
$\int u d u/(1-u^2)$, evaluated in equation
 (\ref{E-basic-integral}), to
write

\begin{eqnarray}
Q= &+& {\pi\over2} 
\nonumber\\
   &+& \Im \left[\int_0^\infty {u du\over1-u^2}
\left\{ { \rho \sqrt{1-u^2} - \sqrt{n^2-u^2}  \over 
  \rho \sqrt{1-u^2} + \sqrt{n^2-u^2} } + 1\right\} \right].
\nonumber\\
&&
\end{eqnarray}

\noindent
This conveniently gets rid of the pole so that the integral is now
unambiguously finite. Indeed

\begin{equation}
Q =+{\pi\over2}+ 2 \rho \Im \left[\int_0^\infty {u du\over1-u^2}
{\sqrt{1-u^2}  \over 
  \rho \sqrt{1-u^2} + \sqrt{n^2-u^2} }\right].
\end{equation}

\noindent
Now we also have to take $n$ to be real, otherwise we step outside
the Balian--Bloch formalism. For now, also take $n>1$, the alternative
case being completely analogous.  The integrand is now imaginary
only over the range $u\in[1,n]$, and we can change variables to
set

\begin{equation}
Q =+{\pi\over2} +  \rho \left[\int_1^{n^2} {d u'\over1-u'}
\Im\left\{  {i\sqrt{u'-1}  \over 
  \sqrt{n^2-u'} + i \rho \sqrt{u'-1} }\right\}\right].
\end{equation}

\noindent
That is 

\begin{equation}
Q =+{\pi\over2}  +  \rho \left[\int_1^{n^2} {d u'\over1-u'}
{\sqrt{n^2-u'} \sqrt{u'-1} \over 
 (n^2-u') + \rho^2(u'-1) }\right].
\end{equation}

\noindent
Equivalently

\begin{equation}
Q =+{\pi\over2} -  \rho \int_0^{n^2-1} {d u''\over u''}
{\sqrt{n^2-1-u''} \sqrt{u''} \over 
 (n^2-1-u'') +\rho^2 u'' }.
\end{equation}

\noindent
Now define $u''=(n^2-1)w$.
 
\begin{equation}
Q =+{\pi\over2} -  \rho \int_0^1 {d w\over w}
{\sqrt{1-w} \sqrt{w} \over 
1+(\rho^2-1) w }.
\end{equation}

\noindent
Note that the refractive index $n$ has now completely disappeared
from the integral. This gives
 
\begin{equation}
Q 
=+{\pi\over2} -  \rho {\pi\over \rho +1} 
=-{\pi\over2} \left[ {\rho-1\over\rho+1}\right]. 
\end{equation}

\noindent
We can re-do the calculation for  $n<1$.  A few intermediate steps
change but the final result is the same.  We finally have our
announced result

\begin{equation}
\xi_{\small out} (\rho, n) = 
{\pi\over4} \left[ {\rho-1\over\rho+1}\right]
={\pi\over4} \left[ {\rho_1-\rho_2\over\rho_1+\rho_2}\right].
\end{equation}

\noindent
Note the remarkable result that this is independent of $n$ for $n$
real. With hindsight, we can see that the acoustic junction
conditions explicitly make reference only to the density of the fluid,
and not to the velocity of sound (refractive index), which might be
viewed as an {\em a posteriori} justification for the absence of
refractive index in the final result. However we know of no simple
physics argument that would justify this, and must rely on the
explicit calculation presented above.

As $\rho\to+\infty$ we recover  Neumann and
Robin boundary conditions while as $\rho\to0$ we recover Dirichlet
boundary conditions. Also note that
on interchanging the two regions, $\rho\to{1/ \rho}$, so
we have

\begin{equation}
\xi_{\small in}(\rho) = 
\xi_{\small out}(1/\rho) = 
-\xi_{\small out}(\rho),
\end{equation}

\noindent
as expected from our earlier discussion [see Eq. (\ref{E-inversion})].

\subsection{Electromagnetic field}

For the electromagnetic field, we can use the analysis presented by
Balian and Bloch in~\cite[pages 273--274]{Balian-Bloch-71} to view the
electromagnetic eigenvalue problem as a combination of vector and
scalar eigenvalue problems.  A standard result is
\\

\noindent 
{\bf Perfect conductor boundary conditions:}
\\ 
($\vec E \times \vec n = 0$ and $\vec B\cdot\vec n = 0$ on the boundary) 

\begin{equation}
\xi = 0.  
\end{equation}

This vanishing of the surface term for perfect conductor boundary
conditions is due to a cancellation between TE and TM modes. (For a
surface of general shape the separation into TE and TM modes is
meaningless; TE and TM modes make sense only in situations of
extremely high symmetry. Nevertheless, sufficiently close to any
conducting surface we may approximate the surface by its tangent
plane---and in this {\em approximation} the decomposition into TE and
TM modes makes sense. The general vector minus scalar decomposition
alluded to above then {\em approximately} reduces to the simpler
scalar plus scalar decomposition for the TE and TM modes.)
\\

\noindent 
{\bf Dielectric junction conditions:}
\\ 
For the case of ultimate interest we are of course interested in {\em
dielectric junction conditions}.  A full appreciation of the (perhaps
unexpected) subtleties involved with dielectric junction conditions
might be gleaned from the fact that even for a plane interface the
situation is sufficiently complicated as to warrant a recent 600 page
technical monograph~\cite{King-Owens-Wu}, and a continuing stream of
research papers (see for instance~\cite{GRV}).

Nevertheless we can make a few general statements on physical grounds
before doing a detailed calculation of $\xi$.  In analogy with the case
of the scalar field, finite-volume effects will distort the density of
states both {\em inside} and {\em outside} the dielectric body
according to the general scheme

\begin{eqnarray}
\sum_{inside} &\sim& V \int { d^3 \vec k\over (2\pi)^3} +
            S \int \xi_{\small in} { d^3 \vec k \over (2\pi)^3 k} + \cdots
\\
\sum_{outside} &\sim& (V_\infty-V) \int { d^3 \vec k\over (2\pi)^3} +
            S \int \xi_{\small out} { d^3 \vec k \over (2\pi)^3 k} + \cdots
\end{eqnarray} 

\noindent
For the case of a dielectric junction, we expect $\xi(\epsilon,\mu)$
to be a function of the permeability and permitivity, and we know,
from first principles, that as $\epsilon\to 1$ and $\mu\to1$ the
dielectric boundary disappears as both media become the same, so we
must have

\begin{equation}
\xi(\epsilon,\mu)\to 0 \qquad {\rm as} \qquad 
\epsilon\to 1 \quad {\rm and} \quad \mu\to 1.
\end{equation}

\noindent
When we turn to including dispersive effects we note that
$\xi(\epsilon,\mu)$ should ultimately be taken to be a function of the
wave-number dependent quantities $\epsilon(k)$, $\mu(k)$. Since we
know that as $k\to \infty$ the dielectric must ultimately mimic
individual atoms embedded in vacuum, we must have

\begin{equation}
\xi(\epsilon(k),\mu(k))\to0 \qquad {\rm as} \qquad  k\to\infty.
\end{equation}

The calculation of  $\xi$ for the electromagnetic field is an easy
exercise given our results for the acoustic problem. We decompose
the electromagnetic field near the approximately plane boundary
into TE and TM modes. In terms of the relative refractive index,
relative permitivity, and relative permeability, the reflection
coefficients (for the outside region) are simply~\cite[pages
83--84]{DeSanto}  (or see~\cite[equations (86.4) and (86.6), page
295]{Landau-Lifshitz}, or~\cite[pages 281--282]{Jackson})\footnote{Be
careful with all the different notations in use.}

\begin{eqnarray}
R^{\small TE}(\epsilon,\mu;u) 
&=&
R^{\small acoustic}(\rho=\mu,n;u)
\nonumber\\
&=& 
{\mu\sqrt{1-u^2} - \sqrt{n^2-u^2}  \over 
 \mu\sqrt{1-u^2} + \sqrt{n^2-u^2} }.
\end{eqnarray}

\begin{eqnarray}
R^{\small TM}(\epsilon,\mu;u) 
&=& 
R^{\small acoustic}(\rho=\epsilon,n;u) 
\nonumber\\
&=& 
{\epsilon\sqrt{1-u^2} - \sqrt{n^2-u^2}  \over 
 \epsilon\sqrt{1-u^2} + \sqrt{n^2-u^2} }.
\end{eqnarray}

\noindent
(Remember that $n=\sqrt{\epsilon\mu}$. Also,
we have defined $n=n_1/n_2$, $\epsilon=\epsilon_1/\epsilon_2$, and
$\mu=\mu_1/\mu_2$)

Thus, applying the previous acoustic results, we get the remarkably
simple formulae:

\begin{equation}
\xi^{\small TE}_{\small out}(\mu) =
{\pi\over4}\left[{\mu-1\over \mu+1}\right] = 
{\pi\over4}\left[{\mu_1-\mu_2\over \mu_1+\mu_2}\right] =
- \xi^{\small TE}_{\small in}(\mu).
\end{equation}

\begin{equation}
\xi^{\small TM}_{\small out}(\epsilon) = 
{\pi\over4}\left[{\epsilon-1\over \epsilon+1}\right] =
{\pi\over4}\left[{\epsilon_1-\epsilon_2\over \epsilon_1+\epsilon_2}\right] = 
- \xi^{\small TM}_{\small in}(\epsilon).
\end{equation}

Note that the result for the TE modes is independent of $\epsilon$,
while that for the TM mode is independent of $\mu$. Since most
typical dielectric materials are magnetically inert, $\mu\approx1$,
the TE contribution is typically much smaller than the TM
contribution.

{\em Consistency check:} Instead of appealing to the identification of
reflection coefficients, we can get the same results directly from the
dielectric boundary conditions.  We know that

\begin{equation}
\vec E^\perp, \quad \epsilon \vec E^n, \quad
\vec H^\perp, \quad {\rm and} \quad \mu \vec H^n,
\end{equation}
must be continuous across the boundary.

If we are dealing with a plane interface, or in the approximation that
we are sufficiently close to a curved interface, specifying the normal
components of the $\vec E$ and $\vec B$ fields is sufficient to
completely determine the electromagnetic field. In terms of these
normal components the junction conditions are simply

{\em TE mode:}
\begin{equation}
\label{E-TE-junction-condition-1}
\mu_1 H_1^n = \mu_2 H_2^n,
\end{equation}

\begin{equation}
\label{E-TE-junction-condition-2}
\partial_n H_1^n = \partial_n H_2^n.
\end{equation}

{\em TM mode:}
\begin{equation}
\label{E-TM-junction-condition-1}
\epsilon_1 E_1^n = \epsilon_2 E_2^n,
\end{equation}

\begin{equation}
\label{E-TM-junction-condition-2}
\partial_n E_1^n = \partial_n E_2^n.
\end{equation}

Applying the formalism derived for the acoustic junction conditions,
the previously quoted results for $\xi$ immediately follow.

\section{The Casimir energy}

\noindent
Including these surface contributions to the density of states,
the total zero-point energy for a dielectric body embedded in a
background dielectric is easily seen to be

\begin{eqnarray}
E_{\small embedded-body}
 &=& 2 
V \int {d^3 \vec k\over(2\pi)^3} {1\over2}\hbar \; \omega_1(k)
\nonumber\\
&+&
2 S \int {d^3 \vec k\over(2\pi)^3} {1\over2} \hbar c
\left[ \bar\xi_{\small in}(\epsilon,\mu)   {\omega_1(k)\over c k} \right] 
\nonumber\\
&+& 
2 (V_\infty-V)\int {d^3 \vec k\over(2\pi)^3} {1\over2}\hbar \; \omega_2(k)
\nonumber\\
&+&
2 S \int {d^3 \vec k\over(2\pi)^3} {1\over2} \hbar c
\left[ \bar\xi_{\small out}(\epsilon,\mu) {\omega_2(k)\over c k}\right] 
\nonumber\\
&+& \cdots
\end{eqnarray}

\noindent
This is just the generalization of equation (\ref{E-embedded-body})
above to include surface effects. The quantity $\bar\xi$ denotes
an {\em average} over TE and TM modes. To calculate the Casimir energy
we now simply subtract the homogeneous dielectric zero-point energy
[equation (\ref{E-Minkowski})] to obtain

\begin{eqnarray}
E_{\small Casimir} 
&=& 
2 V \int {d^3 \vec k\over(2\pi)^3} {1\over2}\hbar
\left[ \omega_1(k) - \omega_2(k) \right]
\nonumber\\
&+&
2 S \int {d^3 \vec k\over(2\pi)^3} {1\over2} \hbar c
\left[ {\bar\xi_{\small in}(\epsilon,\mu)\over n_1} + 
       {\bar\xi_{\small out}(\epsilon,\mu)\over n_2}\right] 
\nonumber\\
&+& \cdots
\end{eqnarray}

\noindent
Even though the surface terms seem to be additive, there is a
``hidden'' minus sign, as we shall see below, due to the fact that
$\xi_{\small in}(\rho) = -\xi_{\small out}(\rho)$.

This is quite enough to give a good qualitative feel for the physics:
the Casimir effect will in general induce a surface tension that
goes as $(\rm{cutoff})^3$.

It is useful to define

\begin{equation}
\Xi(\epsilon_1,\mu_1;\epsilon_2,\mu_2) =
\left[ {\bar\xi_{\small in}(\epsilon,\mu)\over n_1} + 
       {\bar\xi_{\small out}(\epsilon,\mu)\over n_2}\right]
 \end{equation}

\noindent
and so write the Casimir surface tension as

\begin{equation}
\sigma({\rm surface~tension}) =  
\int {d^3 \vec k\over(2\pi)^3} \hbar c \; \;
\Xi(\epsilon_1,\mu_1;\epsilon_2,\mu_2).
\end{equation}

\noindent
{From} our previous results for $\xi$, taking the case of magnetically
inert media for simplicity ($\mu=1$), we see

\begin{equation}
\Xi(n_1,n_2) =
{\pi\over8}\left[- {1\over n_1} + {1\over n_2}\right]
{n_1^2-n_2^2\over n_1^2+n_2^2}.
\end{equation}

\noindent
Here we indeed see that the two surface terms contribute with opposite
signs, largely cancelling each other.  We can factorize this to yield

\begin{equation}
\Xi(n_1,n_2) =
+ {\pi\over8}
{(n_1-n_2)^2 (n_1+n_2)\over n_1 n_2 (n_1^2+n_2^2)}.
\end{equation}

Note that this vanishes as $(n_1-n_2)^2$, with one factor of
$(n_1-n_2)$ coming from the fact that the $\xi_i$ individually tend to
zero as $n_1 \to n_2$ and the second coming from the partial
cancellation discussed above.

\noindent
What does this do to the Casimir energy?

\begin{eqnarray}
E_{\small Casimir} &=& 2 V \int {d^3 \vec k\over(2\pi)^3}
        {1\over2}\hbar c k
           \left[{n_2-n_1\over n_1 n_2}\right]
\nonumber\\
          &+&
           2 S \int {d^3 \vec k\over(2\pi)^3} {1\over2} \hbar c
            {\pi\over8} {(n_1-n_2)^2 (n_1+n_2)\over n_1 n_2 (n_1^2+n_2^2)} 
\nonumber\\
&+& \cdots
\end{eqnarray}

This is our general result for the Casimir energy. We now insert a
momentum dependent refractive index into the above to explictly
evaluate the coefficients. The physical cutoff is provided by the fact
that both refractive indices are known to tend to $1$ at large
momenta.

A naive hard cutoff, following the ideas of Schwinger, simplifies
these expressions considerably. Naive hard cutoffs are of course an
idealization that suppresses much of the physical detail, and are
justified only for order of magnitude estimates and for comparison
with the previous literature where naive hard cutoffs are often the
only extant results. Suppose we take

\begin{equation}
n_1(k) = n_1\; \Theta(K-k) + \Theta(k-K),
\end{equation}

\noindent
and

\begin{equation}
n_2(k) = n_2\; \Theta(K-k) + \Theta(k-K).
\end{equation}

\noindent
(It is an additional gross over-simplification to set the cutoffs for
the two media equal to one another, but it is standard and is the only
way to make connection with previous calculations. Keeping separate
cutoffs for the two media is straightforward but algebraically
somewhat messy.)

The Casimir energy is then given by

\begin{eqnarray}
E_{\small Casimir} 
&=& 
{1\over8\pi^2} V \; \hbar c \;  K^4 
\left[{1\over n_1} - {1\over n_2} \right]
\nonumber\\
&+&  {1\over6\pi^2} S \; \hbar c \; K^3  \; 
\left[ 
{\bar\xi_{\small in}(n_1,n_2)\over n_1} +
{\bar\xi_{\small out}(n_1,n_2)\over n_2} 
\right]
\nonumber\\
&+& \cdots,
\end{eqnarray}

\noindent
while the Casimir surface tension is

\begin{equation}
\sigma =   {1\over6\pi^2} \hbar c \; K^3 \;  
\left[ 
{\bar\xi_{\small in}(n_1,n_2)\over n_1} + 
{\bar\xi_{\small out}(n_1,n_2)\over n_2} 
\right].
\end{equation}

\noindent
Inserting the specific formulae for $\xi$ then yields
 
\begin{equation}
\sigma = +{1\over48\pi} \hbar c 
{(n_1-n_2)^2 (n_1+n_2)\over n_1 n_2 (n_1^2+n_2^2)} 
K^3.
\end{equation}

\noindent
Now particularize to dilute media, by taking $n_1\approx1\approx n_2$. 

\begin{eqnarray}
E_{\small Casimir} 
&\approx& 
{1\over8\pi^2} V \; \hbar c \;  K^4 
\left[ n_2 - n_1 \right]
\nonumber\\
&+&  {1\over48\pi} S \; \hbar c \; K^3  \; 
\left[ (n_1-n_2)^2 \right]
+ \cdots
\end{eqnarray}

\noindent
The volume term here is the dilute medium limit of Schwinger's
result~\cite{Schwinger0}, while the surface area term reproduces the
Milton {\em et al.}  result~\cite{Milton80,Milton95,Milton96}.  There
is an overall normalization difference between this surface term and
the special case calculated by Milton {\em et. al.}, this
normalization difference being attributable to a different choice of
regulator. The critical physics lies in the volume versus surface area
dependence, the power of the cutoff dependence, and the behaviour as a
function of refractive index.

Note that the bulk term is dominant if

\begin{equation}
V \; K \gg S,
\end{equation}

\noindent
that is, for dielectrics with linear dimensions satisfying

\begin{equation}
L \sim (V/S) \gg 1/K = \lambda_0/(2\pi).
\end{equation}

\noindent
For a typical dielectric we estimate $\lambda_0\approx 1,000$
Angstrom, so for dielectrics of this size or greater the Casimir
energy will be dominated by bulk effect.  {\em This is certainly
the case for sonoluminescence where typical bubble radii are of
order $100,000$ Angstroms. } 

For small enough dielectric particles
the surface term will not be negligible in comparison to the volume
term---this is no great surprise to people studying mesoscopic
systems for which the existence of finite volume effects is well
known.

Finally, we mention that for [non-dispersive]  Neumann, Dirichlet,
and Robin boundary conditions the existence of a surface term
contributing to the total Casimir energy has been known for some
time---see for instance~\cite{BVW}.

\section{Discussion}

The main results of this paper are:\\
{}(1) The Casimir energy in a dielectric medium is dominated
by a volume term. Indeed, for a finite-volume of dielectric 1,
embedded in an infinite volume of different dielectric 2,

\begin{eqnarray}
E_{\small Casimir} &=& 2 V \int {d^3 \vec k\over(2\pi)^3} {1\over2} \hbar
\left[ \omega_1(k) - \omega_2(k) \right]
\nonumber\\
&+& 2 S \int {d^3 \vec k\over(2\pi)^3} {1\over2} \hbar c \; 
\Xi(\epsilon_1,\mu_1;\epsilon_2,\mu_2)
\nonumber\\
&+& \cdots,
\end{eqnarray}

\noindent
where the dots represent terms arising from higher-order distortions 
of the density of states due to finite-volume effects.

\noindent
{}(2) If we adopt a simple cutoff model for the dispersion relation, 
the volume term is

\begin{equation}
E^{\small bulk}_{\small Casimir} = {1\over8\pi^2} \; V \; \hbar c \; K^4 \;
\left[{1\over n_1} - {1\over n_2} \right].
\end{equation}

\noindent
This result is completely in agreement with Schwinger's calculation
in~\cite{Schwinger0}, and in disagreement with~\cite{Milton95,Milton96}.

\noindent
{}(3) In addition, there will be a sub-dominant contribution to
the Casimir energy that is proportional to the surface area of the
dielectric. This surface contribution takes the generic form

\begin{equation}
E^{\small surface}_{\small Casimir} = 
+{1\over48\pi} \; S \; \hbar c \; K^3 \;  
\left[ {(n_1-n_2)^2 (n_1+n_2) \over n_1 n_2 (n_1^2+n_2^2)} \right].
\end{equation}

\noindent
This term is sub-dominant provided

\begin{equation}
V/S \gg 1/K = \lambda_0/(2\pi).
\end{equation}

\noindent
{}(4) In general, we can expect these to be the first two terms of
a more general expansion that includes terms proportional to various
geometrical invariants of the body. By analogy with the situation
for non-dispersive Dirichlet,  Neumann, and Robin boundary
conditions~\cite{Balian-Bloch} we expect the next term to be
proportional to the trace of the extrinsic curvature integrated
over the surface of the body.

\noindent
{}(5) The analysis of the present paper has been limited to situations
of real refractive index (loss-free insulating dielectrics).
Generalizing to lossy conducting media is clearly of interest but will
require a careful re-assessment of the entire formalism.

\acknowledgements

This work was supported by the U.S. Department of Energy.  Part
of this work was carried out at the Laboratory for Space Astrophysics
and Fundamental Physics (LAEFF, Madrid), and the authors wish to both
acknowledge the hospitality shown, and to acknowledge partial support
by the Spanish Ministry of Education and Culture. Part of this work
was carried out at Los Alamos National Laboratory, and M.V. wishes to
acknowledge the hospitality shown.

The authors wish to thank Juan P\'erez--Mercader (LAEFF), Salman Habib
(Los Alamos), and Carl Carlson (William \& Mary) for many useful
discussions and for carefully reading the manuscript.


\end{document}